\documentstyle[aps,prl,epsf]{revtex}
\bibstyle{unsrt}

\begin{document}
\draft

\twocolumn[\hsize\textwidth\columnwidth\hsize\csname
@twocolumnfalse\endcsname


\title{Quasi-exactly solvable quartic potential}

\author{Carl M. Bender}
\address{Department of Physics, Washington University, St. Louis, MO 63130, USA}

\author{Stefan Boettcher}
\address{Center for Nonlinear Studies, Los Alamos National Laboratory, Los
Alamos, NM 87545, USA\\
and\\
Center for Theoretical Studies of Physical Systems, Clark Atlanta University,
Atlanta, GA 30314, USA}

\date{\today}
\maketitle

\begin{abstract}
A new two-parameter family of quasi-exactly solvable quartic polynomial
potentials $V(x)=-x^4+2iax^3+(a^2-2b)x^2+2i(ab-J)x$ is introduced. Until now, it
was believed that the lowest-degree one-dimensional quasi-exactly solvable
polynomial potential is sextic. This belief is based on the assumption that the
Hamiltonian must be Hermitian. However, it has recently been discovered that
there are huge classes of non-Hermitian, ${\cal PT}$-symmetric Hamiltonians
whose spectra are real, discrete, and bounded below. Replacing Hermiticity by
the weaker condition of ${\cal PT}$ symmetry allows for new kinds of
quasi-exactly solvable theories. The spectra of this family of quartic
potentials discussed here are also real, discrete, and bounded below, and the
quasi-exact portion of the spectra consists of the lowest $J$ eigenvalues. These
eigenvalues are the roots of a $J$th-degree polynomial.
\end{abstract}
]

Quantum-mechanical potentials are said to be {\em quasi-exactly solvable} (QES)
if a finite portion of the energy spectrum and associated eigenfunctions can be
found exactly and in closed form \cite{Ush}. QES potentials depend on a
parameter $J$; for positive integer values of $J$ one can find exactly the first
$J$ eigenvalues and eigenfunctions, typically of a given parity. QES systems can
be classified using an algebraic approach in which the Hamiltonian is expressed
in terms of the generators of a Lie algebra \cite{Tur,Tur1,ST,GKO}. This
approach generalizes the dynamical-symmetry analysis of {\em exactly solvable}
quantum-mechanical systems, whose {\em entire} spectrum may be found in closed
form by algebraic means \cite{Iac}.

An especially simple and well known example of a QES potential \cite{BD} is
\begin{eqnarray}
V(x)=x^6-(4J-1)x^2.
\label{e1}
\end{eqnarray}
The Schr\"odinger equation, $-\psi''(x)+[V(x)-E]\psi(x)=0$,
has $J$ even-parity solutions of the form
\begin{eqnarray}
\psi(x)=e^{-x^4/4}\sum_{k=0}^{J-1}c_k x^{2k}.
\label{e2}
\end{eqnarray}
The coefficients $c_k$ for $0\leq k\leq J-1$ satisfy the recursion relation
\begin{eqnarray}
4(J-k)c_{k-1}+Ec_k+2(k+1)(2k+1)c_{k+1}=0,
\label{e3}
\end{eqnarray}
where we define $c_{-1}=c_{J}=0$. The simultaneous linear equations (\ref{e3})
have a nontrivial solution for $c_0,\,c_1,\,...,\,c_{J-1}$ if the determinant of
the coefficients vanishes. For each integer $J$ this determinant is a polynomial
of degree $J$ in the variable $E$. The roots of this polynomial are all real
and are the $J$ quasi-exact energy eigenvalues of the potential (\ref{e1}).

The lowest-degree one-dimensional QES polynomial potential that is discussed in
the literature is sextic. However, in this paper we introduce an entirely new
two-parameter class of QES {\it quartic} polynomial potentials. The spectra of
this family of potentials are real, discrete, and bounded below. Like the
eigenvalues of the potential (\ref{e1}), the lowest $J$ eigenvalues of these
potentials are the roots of a polynomial of degree $J$.

The potentials introduced here have not been discovered so far because they are
associated with non-Hermitian Hamiltonians. Recently, it has been found that
there are large classes of non-Hermitian Hamiltonians whose spectra are real and
bounded below \cite{A,B}. Although they are non-Hermitian, these Hamiltonians
exhibit the weaker symmetry of ${\cal PT}$ invariance. A class of these
Hamiltonians,
\begin{eqnarray}
H=p^2-(ix)^N\quad(N\geq2),
\label{e4}
\end{eqnarray}
was studied in Ref.~\cite{A}. The special case $N=4$ corresponds to the
Hamiltonian
\begin{eqnarray}
H=p^2-x^4.
\label{e5}
\end{eqnarray}
It is not at all obvious that this Hamiltonian has a positive, real, discrete
spectrum. To verify this property, we must continue analytically the
Schr\"odinger equation eigenvalue problem associated with $H$ in (\ref{e4}) from
the conventional harmonic oscillator ($N=2$) to the case $N=4$. In doing so, the
boundary conditions at $|x|=\infty$ rotate into the complex $x$ plane. At $N=4$
the boundary conditions on the wave function $\psi(x)$ read
\begin{eqnarray}
\lim_{|x|\to\infty}\psi(x)=0,
\label{e6}
\end{eqnarray}
where the limit $x\to\infty$ is taken inside two wedges bounded by the Stokes'
lines of the differential equation. The right wedge is bounded by the Stokes'
lines at $0^\circ$ and $-60^\circ$ and the left wedge is bounded by the Stokes'
lines at $-120^\circ$ and $-180^\circ$. The leading asymptotic behavior of the
wave function is given by
\begin{eqnarray}
\psi(x)\sim e^{-ix^3/3}\quad(|x|\to\infty).
\label{e7}
\end{eqnarray}
It is easy to see that the asymptotic conditions in (\ref{e6}) are
satisfied by $\psi(x)$. A complete discussion of the analytic
continuation of eigenvalue problems into the complex plane is given in
Ref.~\cite{C}. Note that for all values of $N$ between 2 and 4, the Hamiltonian
(\ref{e4}) is not symmetric under parity. This parity noninvariance persists
even at $N=4$; eigenfunctions $\psi(x)$ of (\ref{e5}) are not symmetric (or
antisymmetric) under the replacement $x\to-x$.

In this paper we generalize the Hamiltonian (\ref{e5}) to the two-parameter
class
\begin{eqnarray}
H=p^2-x^4+2iax^3+(a^2-2b)x^2+2i(ab-J)x,
\label{e8}
\end{eqnarray}
where $a$ and $b$ are real and $J$ is a positive integer. The wave function
$\psi(x)$ satisfies the boundary conditions (\ref{e6}) and the differential
equation
\begin{eqnarray}
E\psi(x)=-\psi''(x)&+&\big[-x^4+2iax^3+(a^2-2b)x^2\nonumber\\
&&~~+2i(ab-J)x\big]\psi(x).
\label{e9}
\end{eqnarray}

We obtain the QES portion of the spectrum of $H$ in (\ref{e8}) as follows. We
make the {\it ansatz}
\begin{eqnarray}
\psi(x)=e^{-ix^3/3-ax^2/2-ibx}P_{J-1}(x),
\label{e10}
\end{eqnarray}
where
\begin{eqnarray}
P_{J-1}(x)=x^{J-1}+\sum_{k=0}^{J-2}c_k x^k
\label{e11}
\end{eqnarray}
is a polynomial in $x$ of degree $J-1$. Substituting $\psi(x)$ in (\ref{e10})
into the differential equation (\ref{e9}), dividing off the exponential in
(\ref{e10}), and collecting powers of $x$, we obtain a polynomial in $x$ of
degree $J-1$. Setting the coefficients of $x^k$ ($1\leq k\leq J-1$) to $0$ gives
a system of $J-1$ simultaneous linear equations for the coefficients $c_k$
($0\leq k \leq J-2$). We solve these equations and substitute the values of
$c_k$ into the coefficient of $x^0$. This gives a polynomial $Q_J(E)$ of degree
$J$ in the energy eigenvalue $E$. The coefficients of this polynomial are
functions of the parameters $a$ and $b$ of the Hamiltonian $H$ in (\ref{e8}).
The first five of these polynomials are
\begin{eqnarray}
Q_1 &=& E -b^2 -a,\nonumber\\
Q_2 &=& E^2 -(2b^2+4a)E+b^4+4ab^2-4b+3a^2,\nonumber\\
Q_3 &=& E^3 -(3b^2+9a)E^2+(3b^4+18ab^2-16b+23a^2)E\nonumber\\
&&\quad -b^6-9ab^4+16b^3-23a^2b^2+48ab-15a^3-16,\nonumber\\
Q_4 &=& E^4 -(4b^2+16a)E^3+(6b^4+48ab^2-40b\nonumber\\
&&\quad +86a^2)E^2 +(-4b^6-48ab^4+80b^3-172a^2b^2\nonumber\\
&&\quad +320ab-176a^3-96)E+b^8+16ab^6-40b^5\nonumber\\
&&\quad +86a^2b^4-320ab^3+176a^3b^2+240b^2\nonumber\\
&&\quad -568a^2b+105a^4+384a,
\nonumber\\
Q_5 &=& E^5-(5b^2+25a)E^4+(10b^4+100ab^2-80b\nonumber\\
&&\quad +230a^2)E^3+(-10b^6-150ab^4+240b^3\nonumber\\
&&\quad -690a^2b^2+1200ab-950a^3-336)E^2+(5b^8\nonumber\\
&&\quad +100ab^6-240b^5+690a^2b^4-2400ab^3\nonumber\\
&&\quad +1900a^3b^2+1696b^2-5488a^2b+1689a^4\nonumber\\
&&\quad +3360a)E-b^{10}-25ab^8+80b^7-230a^2b^6\nonumber\\
&&\quad +1200ab^5-950a^3b^4-1360b^4+5488a^2b^3\nonumber\\
&&\quad -1689a^4b^2-8480ab^2+7440a^3b+3072b\nonumber\\
&&\quad -945a^5-7632a^2.
\label{e12}
\end{eqnarray}
The roots of $Q_J(E)$ are the QES portion of the spectrum of $H$.

The polynomials $Q_J(E)$ simplify dramatically if we substitute
\begin{eqnarray}
E=F+b^2+Ja
\label{e13}
\end{eqnarray}
and
\begin{eqnarray}
K=4b+a^2.
\label{e14}
\end{eqnarray}
The new polynomials have the form
\begin{eqnarray}
Q_1 &=& F,\nonumber\\
Q_2 &=& F^2 -K,\nonumber\\
Q_3 &=& F^3 -4KF-16,\nonumber\\
Q_4 &=& F^4 -10KF^2 -96F+9K^2,\nonumber\\
Q_5 &=& F^5-20KF^3 -336F^2+64K^2F+768K,\nonumber\\
Q_6 &=& F^6-35KF^4 -896F^3+259K^2F^2+7040KF\nonumber\\
&&\quad  -225K^3+25600,\nonumber\\
Q_7 &=& F^7-56KF^5 -2016F^4+784K^2F^3+35712KF^2\nonumber\\
&&\quad -2304K^3F+288000F-55296K^2,
\nonumber\\
Q_8 &=& F^8-84KF^6 -4032F^5+1974K^2F^4
\nonumber\\
&&\quad +132480KF^3-12916K^3F^2+1760256F^2
\nonumber\\
&&\quad -681408K^2F+11025K^4-6322176K.
\label{e15}
\end{eqnarray}
The roots of these polynomials are all real so long as $K\geq K_{\rm critical}$,
where $K_{\rm critical}$ is a function of $J$. The first few values of 
$K_{\rm critical}$ are listed in Table \ref{table1}. At $K=K_{\rm critical}$ the
lowest two eigenvalues become degenerate and when $K<K_{\rm critical}$ some of
the eigenvalues of the QES spectrum are complex. Thus, the QES spectrum is
entirely real above a parabolic shaped region in the $(a,b)$ plane bounded by
the curve $a^2+4b=K_{\rm critical}$.

\begin{figure}
\epsfxsize=2.2truein
\hskip 0.15truein\epsffile{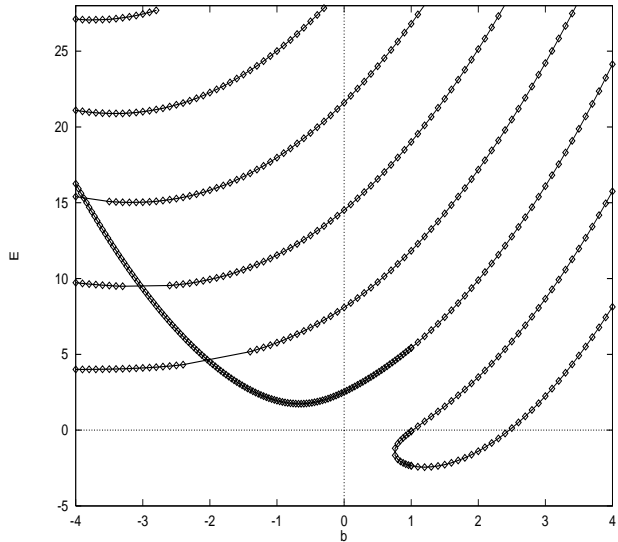}
\caption{
\narrowtext
The spectrum for the QES Hamiltonian (9) plotted as a function of
$b$ for the case $J=3$ and $a=0$. For $b>{3\over4}$ (corresponding to the
critical value $K_{\rm critical}=3$) the QES eigenvalues are real and are the
three lowest eigenvalues of the spectrum. When $b$ goes below $3\over4$, two of
the QES eigenvalues become complex and the third moves into the midst of the
non-QES spectrum. We believe that the non-QES spectrum is entirely real
throughout the $(a,b)$ plane.}
\label{fig1}
\end{figure}

Extensive numerical calculations lead us to believe that the non-QES spectrum is
entirely real throughout the $(a,b)$ plane and that when $K>K_{\rm critical}$ 
the eigenvalues of the QES spectrum lie below the eigenvalues of the non-QES
spectrum. However, as we enter the region $K<K_{\rm critical}$ some of the
eigenvalues of the QES spectrum pair off and become complex. Other eigenvalues
of the QES spectrum may cross above the eigenvalues of the non-QES spectrum.
In Fig.~\ref{fig1} we illustrate the case $J=3$ and $a=0$. Note that for
$b>{3\over4}$ the QES eigenvalues are three lowest eigenvalues of the spectrum.
When $b$ goes below $3\over4$, two of the QES eigenvalues become complex and the
third moves into the midst of the non-QES spectrum.

The standard way to understand QES theories is to demonstrate that the
Hamiltonian can be expressed in

\begin{table}
\caption[t1]{Sequence of critical values for $K_{\rm critical}$ and
$F_{\rm critical}$.}
\begin{tabular}{cdd|cdd}
$J$ & $K_{\rm critical}$& $F_{\rm critical}$& $J$ & $K_{\rm critical}$&
$F_{\rm critical}$\\ 
\tableline
 2&  0.0     &  0.0     & 16 & 25.0526 &  -61.3470\\
 3&  3.0     & -2.0     & 17 & 26.3475 &  -67.3089\\
 4&  5.47086 & -4.71894 & 18 & 27.6149 &  -73.4116\\
 5&  7.65570 & -7.93982 & 19 & 28.8569 &  -79.6490\\
 6&  9.65184 & -11.5572 & 20 & 30.0754 &  -86.0158\\
 7& 11.5104  & -15.5070 & 21 & 31.2721 &  -92.5072\\
 8& 13.2625  & -19.7459 & 22 & 32.4485 &  -99.1187\\
 9& 14.9287  & -24.2419 & 23 & 33.6058 & -105.846\\ 
10& 16.5235  & -28.9706 & 24 & 34.7453 & -112.686\\ 
11& 18.0576  & -33.9126 & 25 & 35.8679 & -119.635\\ 
12& 19.5392  & -39.0521 & 26 & 36.9747 & -126.689\\ 
13& 20.9747  & -44.3758 & 27 & 38.0665 & -133.846\\ 
14& 22.3695  & -49.8725 & 28 & 39.1439 & -141.103\\
15& 23.7276  & -55.5323 & 29 & 40.2078 & -148.458\\
\end{tabular}
\label{table1}
\end{table}

\noindent
terms of generators of a Lie algebra. Following Turbiner
\cite{Tur1}, we use the generators of a finite dimensional
representation of the $SL(2,Q)$ with spin $J$. The three generators have the
form
\begin{eqnarray}
{\cal J}^+&=&x^2{d\over dx}-(J-1)x,\quad {\cal J}^0=x{d\over dx}-{J-1\over2},
\nonumber\\
{\cal J}^-&=&{d\over dx}.
\label{e16}
\end{eqnarray}
If we apply the Hamiltonian $H$ in (\ref{e8}) to $\psi(x)$ in (\ref{e10})
and divide of the exponential we obtain an operator $h$ acting on the
polynomial $P_{J-1}(x)$; $h$ has the form
\begin{eqnarray}
h&=&-{d\over dx^2}+(2ix^2+2ax+2ib){d\over dx}\nonumber\\
&&-[2i(J-1)x-b^2-a].
\label{e17}
\end{eqnarray}
Hence, in terms of the generators of the Lie algebra, we have
\begin{eqnarray}
h=-({\cal J}^-)^2 +2i{\cal J}^++2a{\cal J}^0+2ib{\cal J}^- +b^2+aJ.
\label{e18}
\end{eqnarray}
This algebraic structure possesses ${\cal PT}$ symmetry
and has {\it real} eigenvalues.

We thank the U.S. Department of Energy for financial support.

\end{document}